# Web to Semantic Web & Role of Ontology

Zeeshan Ahmed, Detlef Gerhard
Mechanical Engineering Informatics and Virtual Product Development Division, Vienna University of Technology, Getreidemarkt 9/307 1060 Vienna, Austria
004315880130726
{zeeshan.ahmed, detlef.gerhard}@tuwien.ac.at

*Abstract*— **In this research paper we are briefly presenting current major web problems and introducing semantic web technologies with the claim of solving existing web's problems. Furthermore we are describing Ontology as the main building block of semantic web and focusing on its contributions to semantic web progress and current limitations.**

## I. INTRODUCTION

DATA in the form of audio, video, text and image documents is published on web using Hyper Text Markup Language (HTML), considered to be poor in defining and formalizing the meaning of the context [11]. HTML based documents are formatted in way that these can not be processed because these are only available in readable format. This format deficiency becomes the major cause of some semantic based problems and the need of some other approach which will publish data over the web in not only the readable but also in process able format. If data will be available in both read and process able formats, then it will improve the process of search, extraction and maintenance of data over the web.

Currently two kinds of search techniques exists; Full Text Search (FTS) and Unambiguous Search (US). FTS processes natural language based queries to retrieve information like Google [4] where as US is based on data whose semantic is already defined in the system like Reunion [6]. The probability of getting more concrete and optimized results from US is high as compared to FTS because there is no such mechanism exists which can extract the semantics from full text search query and then look for knowledge based information. To full fill this current need and implement the new idea of data formation, Semantic Web was introduced [5].

Semantic Web is an advanced version of existing web which claims to be a solution toward currently faced web problem of formatting data in machine process able format to take advantage in information retrieval. Semantic web is a mesh of data representing meanings through connectivity, expressing multiple view points and use business rules logic is making models to share information across applications. The goal of semantic web is to structure the meaningful contents of unstructured data and involve knowledge management in making more advanced knowledge modeled management systems [5]. Currently semantic web is solely depending upon a single and main building block Ontology, briefly discussed in section II.

## II. ONTOLOGY

Ontology is the explicit and abstract modeled representation of already defined finite sets of terms and concepts, involved in knowledge engineering, knowledge management and intelligent information integration [9]. Currently, Ontology is divided into three categories .i.e., Natural Language Ontology (NLO), Domain Ontology (DO) and Ontology Instance (OI). NLO is the relationship between generated lexical tokens of statements based on natural language, DO is the knowledge of a particular domain and OI is the automatically generated web page behaves like an object [8].

Ontology development is an iterative process based on seven activities as shown in Fig.1. Ontology development process starts with determining the scope by gathering and listing of terms based unstructured relevant information. It includes information in the form of classes as nouns, properties as verbs containing values, constraints and relationship with the other properties based on classes and their sub classes. Terms are organized in a top down or bottom up class and sub class hierarchy called Taxonomy Hierarchy, works in a way that the instance of the child class can behave as the instance of parent class and the property of the parent class can also be applied to instance of child class. Size of Ontology varies with respect to the number of classes and instances; if the number of classes and instances will increase then the size of ontology will increase.

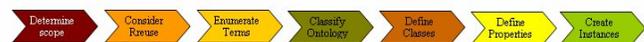

Fig. 1 Ontology development activities [3]

Ontology can be created manually from scratch by extracting information from web and merging already existing ontologies in to new ontologies. Ontology development process with respect to implementation point of view depends on some currently available ontology supported languages XML, RDF and OWL.

*A. XML*

XML (eXtensible Markup Language) [10] is one of the fundamental contributions towards middleware technologies. It is a markup Meta language which allows sharing of information between different applications through markup, structure and transformation. As the major contribution towards semantic web XML provides syntax serialization and abbreviation for data modeling using Data Type Definitions [11] but XML schema is restricted and can only be used for the structured documents because it does not provide semantic, arbitrary naming and structuring of elements

*B. RDF*

RDF (Resource Description Framework) [1] is a URL based syntax data representation which provides a secure and reliable mechanism for metadata exchange between web applications. RDF processes Meta data by making abstract data models based on three object types .i.e., Resource, Property and Statement. Resource is an expression, Property is an attribute describing resource and Statement is a resource having some properties and values. RDF uses three containers .i.e., Object Bag, Sequence and Alternative to arrange available and alternative values in an order [2]. RDF is more useful than XML because it provides independent syntax serialization and abbreviation for data modeling, syntax reification and semantic based features like domain independency, vocabulary and privileges in defining terminologies used in schema language but still RDF modeling mechanism is insufficient in expressing various logical statements [11].

*C. OWL*

OWL (Web Ontology Language) [12] derived from American DARPA Agent Markup Language (DAML) and based on ontology, inference and European Ontology Interchange Language (OIL) [7]. OWL claims to be an extension in RDF in expressing logical statements because it not only describes classes and properties but also provides the concept of namespace, import, cardinality relationship between the classes and enumerated classes. Right now OWL have some limitations like only one Namespace per project is allowed, Import is not currently supported, no database backend and Multi-User support and a few OWL Language features are missing [3].

### III. LIMITATION IN ONTOLOGY

Where ontology is contributing in the progress of semantic web there it also has some limitations.

- Ontology makes the abstract model of a particular domain based on set of data and structures but lacks in defining the boundaries of model.
- Size of Ontology varies with respect to the number of classes and instances; if the number of instances increased to large extent then it becomes very hard to manage manually and currently there is no as such mechanism exists to manage automatically.
- Manual Ontology generation process sometime becomes very complex and time consuming especially while dealing with the large amount of data and to support the process of semantic enrichment reengineering for the building of web consisting of metadata depends on the proliferation of ontologies and relational metadata and requires high production of metadata at high speed and low cost, which is currently also not available.

### IV. CONCLUSION

In this paper we have discussed semantic based information retrieval problem as one of the major cause to improve the concept of web in to semantic web. We have also briefly discussed the contributions, development process, technologies and limitations of Ontology, as an important building block of Semantic Web. This short level conducted research work resulted with the information that the goal of semantic web, to structure the meaningful contents of web to take advantage in information retrieval process and to involve knowledge management in making some more advanced knowledge modeled management systems is not currently achieved. Major contributions are needed in Ontology development process by improving Ontology supported languages RDF and OWL.